\documentclass[aps, prd, twocolumn, superscriptaddress, showkeys, byrevtex groupedaddress]{revtex4}
\usepackage{scalerel}
\usepackage{graphicx}  
\usepackage{dcolumn}   
\usepackage{bm}        
\usepackage{amssymb}   
\usepackage{lipsum}
\usepackage{lineno}
\usepackage{url}
\usepackage{amsmath,array}
\usepackage{natbib}
\usepackage{txfonts}
\usepackage[normalem]{ulem}
\usepackage{soul}
\usepackage{float}
\usepackage{morefloats}
\usepackage{booktabs}
\usepackage{epsfig}
\usepackage{eucal}
\usepackage{hyperref}
\hypersetup{
	colorlinks=true,
	linkcolor=blue,
	filecolor=magneta,      
	urlcolor=blue,
}
\usepackage{tikz}
\usetikzlibrary{svg.path}
\definecolor{orcidlogocol}{HTML}{A6CE39}
\tikzset{
    orcidlogo/.pic={
        \fill[orcidlogocol] svg{M256,128c0,70.7-57.3,128-128,128C57.3,256,0,198.7,0,128C0,57.3,57.3,0,128,0C198.7,0,256,57.3,256,128z};
        \fill[white] svg{M86.3,186.2H70.9V79.1h15.4v48.4V186.2z}
        svg{M108.9,79.1h41.6c39.6,0,57,28.3,57,53.6c0,27.5-21.5,53.6-56.8,53.6h-41.8V79.1z M124.3,172.4h24.5c34.9,0,42.9-26.5,42.9-39.7c0-21.5-13.7-39.7-43.7-39.7h-23.7V172.4z}
        svg{M88.7,56.8c0,5.5-4.5,10.1-10.1,10.1c-5.6,0-10.1-4.6-10.1-10.1c0-5.6,4.5-10.1,10.1-10.1C84.2,46.7,88.7,51.3,88.7,56.8z};
    }
}
\newcommand\orcidicon[1]{\href{https://orcid.org/#1}{\mbox{\scalerel*{
                \begin{tikzpicture}[yscale=-1,transform shape]
                \pic{orcidlogo};
                \end{tikzpicture}
            }{|}}}}
\begin{document}
\title{Viscosity in cosmic fluids}
\author{Pravin Kumar Natwariya \orcidicon{0000-0001-9072-8430}\,}
\email{pvn.sps@gmail.com}
\affiliation{%
    Theoretical Physics Division,Physical Research Laboratory, Ahmedabad 380 009 Gujarat, India}
\affiliation{%
    Department of Physics, Indian Institute of Technology Gandhinagar, Palaj, Gandhinagar 382 355 Gujartat, India}  
\author{Jitesh R Bhatt}
\email{jeet@prl.res.in}
\affiliation{
	Theoretical Physics Division,Physical Research Laboratory, Ahmedabad 380 009 Gujarat, India}
\author{Arun Kumar Pandey}
\email{arunp77@gmail.com}
\affiliation{%
    Department of Physics and Astrophysics, University of Delhi, Delhi 110 007, India}
\date{August 10, 2020}
\begin{abstract}
    {\centering
        \bf Abstract\par
    }
   The effective theory of large-scale structure formation based on $\Lambda$CDM paradigm predicts finite dissipative effects in the resulting fluid equations. In this work, we study how viscous effect that could arise if one includes self-interaction among the dark-matter particles combines with the effective theory. It is shown that these two possible sources of dissipation can operate together in a cosmic fluid and the interplay between them can play an important role in determining dynamics of the cosmic fluid. In particular, we demonstrate that the viscosity coefficient due to self-interaction is added inversely with the viscosity calculated using  effective theory of  $\Lambda$CDM model. Thus the larger viscosity has less significant contribution in the effective viscosity. Using the known bounds on $\,\sigma/m$ for self-interacting dark-matter, where $\,\sigma\,$ and $m$ are the cross-section and mass of the dark-matter particles respectively, we discuss role of the effective viscosity in various cosmological scenarios.

\end{abstract}
\keywords{Viscosity, Boltzmann equation, $\Lambda$CDM, SIDM}
\maketitle
\section{Introduction}
In order to study large scales structures in the Universe, there are two important length-scales: one is comoving Hubble scale $\mathcal{H}^{-1}$ and the another is the non-linear scale $k_{\rm NL}^{-1}$. Here, $k_{\rm NL}^{-1}$ describes the scales at which gravitational collapse takes place; it is typically considered to be of the order of the size of a Galactic cluster, i.e., $\sim $ a few Mpc. The Universe is homogeneous at a scale of $\sim200$~Mpc, and there are roughly $15^3$ homogeneous patches within the Hubble volume. The dynamics of the perturbations can be analyzed in terms of a parameter $\epsilon_{k} =k_{\rm NL}/k$, where $k$ is the inverse length scale. The hierarchy between these two scales is quantified by the parameter  $\epsilon_{k}\gg 1$  which is responsible for the success of linear perturbation theory in describing the observed large scale structures (LSS) (for a recent review see \cite{Bullock:2017B} and also \cite{Baumann:2010tm}). The dark energy (cosmological constant $\Lambda$) plus cold dark-matter (CDM) model, 
(i.e. $\Lambda$CDM) is highly successful in predicting the large scale structure of the Universe. The model is consistent with observations from the length scales typically of the order of $\sim$ 1 Mpc (i.e., intergalactic scale) to the scale of the horizon ($\sim$ 15000 Mpc) \cite{Bullock:2017B}. In this model, structure formation in the dark-matter ( DM ) sector occurs more rapidly than the baryonic matter. The structure formation in the dark sector provides a gravitational potential for the baryonic matter and hence gives the information about the distribution of visible matter in the Universe. Although this model provides extensive agreement with the large scale structure and Cosmic Microwave Background (CMB) radiation observations, it faces difficulty at small length scale ($\lesssim$ 1Mpc). These  problems include `missing satellite problem' \cite{Moore:1999nt, Klypin:1999uc} (prediction of too many dwarf galaxies within the viral radius of the Milky Way from the N-body simulations than observed), the `cusp-core problem' \cite{block:2010wj} (Observations show nearly constant dark-matter density in the inner parts of galaxies, but simulations show a steeper density behavior) and the `too-big-to-fail problem' \cite{Boylan:2011sm, Boylan:2012bs} (from simulations it is not possible to explain the dynamics of the massive satellites in the Milky Way galaxy). Especially, these problems become more evident in studying the galaxy rotation curve \cite{McGaugh2001,Bullock:2017B}.

There have been attempts to address some of these issues within $\Lambda$CDM and also by modifying the $\Lambda$CDM model (see the review \cite{Bullock:2017B} and references therein).  One of the exciting proposals to resolve the issues related to the small scales is by introducing self-interaction between dark-matter particles. Such models are called self-interacting dark-matter (SIDM) models. In these models typical mean free path of dark-matter particle is taken to be in the range of  1 kpc to 1 Mpc, proposed as a remedy for tension between observations and numerical simulations at the scale of a few Mpc ($\epsilon_{k}\ll 1$) \cite{Spergel:2000DSP, Tulin:2017Y, Kaplinghat:2016TY}. Inclusion of interaction can introduce dissipation in the dark-matter fluid, and one can define coefficients of bulk and shear viscosities \cite {Atreya:2018bm}. This small scale physics can affect the large scale behavior of the Universe- it has been shown that the viscous effect can lead to an accelerated expansion of the Universe \cite{Atreya:2018bm,PADMANABHAN:1987, Mohan:2017SM, Floerchinger:2015TW,Das:2012B,Gagnon:2011J,Brevik2017}. Further, the dissipative dynamics of dark-matter can resolve the tension between Planck CMB and LSS observations \cite{Anand:2017}. In other scenarios, viscous cosmology can also be used for constraint the neutrino mass \cite{Anand:2018PASP}. It also explains the cosmic chronometer and type Ia supernova data \cite{Kleidis:2011S, Atreya:2019JA}. As well, dissipations can play a role at suppressing the growth of density perturbations and delaying the nonlinearities in the Universe \cite{Barbosa:2017}. The dissipative effect may arise due to dark-matter-baryon interaction also. Recently a systematical inclusion of baryon-DM interaction has been incorporated in the Boltzmann-Fokker-Planck formalism \cite{Ali:2019, Jacques:1986lj}. It ought to be noted that the baryon-DM interaction has also been considered in the literature to explain 21-cm line \cite{Munoz:2015ed, Barkana:2018or, Bhatt:2019NNP, Bhatt:2019mn}.  The damping of the gravitational waves in the viscous fluid can be used to constrain the mean free path and the DM mass \cite{Lu:2018smr, Goswami:2017CMP}. In this work we critically examine the role of the viscosity that arises due to self-interaction among dark-matter particles.

Before we proceed further, it is important to note that the dissipative effects may arise even for the case of cold-collisionless dark-matter (CCDM) in the presence of self-gravity. In Ref. \cite{Baumann:2010tm} the effective fluid theory of the long-wavelength Universe was obtained by integrating out the short-wavelength perturbations. The effective fluid behaves as a viscous medium coupled to gravity. Here the short-wavelength contributes to the viscous stress tensor of the DM fluid, which depends on the gravitational potential. The effective fluid description of CCDM  is based on the truncation of the Boltzmann hierarchy \cite{Ma:1995E, Baumann:2010tm}. This stress tensor can potentially change the bias parameters in the galaxy bispectrum \cite{Eggemeier2019}. The perturbations contributing to the background in the effective viscous fluid may affect the baryon acoustic oscillation \cite{Baumann:2010tm, Eingorn2019}. If the self-interaction among dark-matter particle is turned on it can change the physics  described in Ref. \cite{Baumann:2010tm} . Thus to incorporate effect of the self-interaction, in the present work, we consider the Boltzmann kinetic equation in the relaxation time approximation to obtain the effective fluid description  for the dark-matter particles. We consider two relaxation times in our scheme: first relaxation time which is inspired by the effective fluid considered in case of CCDM \cite{Baumann:2010tm} and the second relaxation time is based on the cross-section for SIDM \cite{Kaplinghat:2016TY,Atreya:2018bm}. In order to estimate the relaxation time for the interaction among dark-matter particles, we take `SIDM halo model' described in Ref. \cite{Kaplinghat:2016TY}. For the relaxation time due to collision  one writes  $\tau_{si}=1/(n\langle \sigma v_c \rangle)$. The average scattering rate per particle times the halo age can be written as: $(\langle \sigma v_c \rangle/ m)\, \rho\, t_{age} \sim 1$, where $\rho=m n$ with $n$ and $m$ respectively denote number density and mass of dark-matter particles. This expression in Ref. \cite{Kaplinghat:2016TY} used to obtain bounds on $\sigma/m$. For the present work, we take $t_{age} \sim \tau_{si}$. Thus one can allow for more than one sources of viscosity in dark-matter fluid. In such a situation, different viscosities can combine in a particular way. For example, in quark-gluon plasma shear viscosity $\eta_A$ due to turbulence and kinetic viscosity $\eta_c$ combined to give effective shear viscosity  $\eta_{\mathit {eff}}$ as $1/\eta_{\mathit {eff}}=1/\eta_A+1/\eta_c$ \cite{Asakawa:2007sb}. Using the relaxation times, we show that the two different viscosity sources combine in the above way. We believe that this additional contribution to the viscosity can significantly alter the dynamics of the dark-matter fluid and provide useful insight into long-wavelength dynamics of the dark-matter fluid.

This work is divided into following sections: Section [\ref{sect:2}], contains the fluid approximation for the collisionless cold dark-matter in the presence of self-gravity; in section [\ref{sect:3}], we have calculated the relaxation times for collisionless cold dark-matter and self-interacting dark-matter; Spatial perturbation in the Maxwell-Boltzmann (MB) distribution of dark-matter fluid is discussed in section [\ref{sect:4}]; in section [\ref{sect:5}], we get the shear and bulk viscosity for cosmic fluid. Finally, we have given results obtained in present work and a brief conclusion in sections [\ref{sect:6}]. All Latin indexes in the manuscript represents the spatial indices. 


\section{Fluid approximation for CCDM}\label{sect:2}
In this section, we consider identical, non-relativistic, collisionless cold dark-matter particles, coupled gravitationally with each other. Dynamics of phase-space distribution of the particles can be described by Boltzmann Equation \cite{Ma:1995E}
\begin{alignat}{2}
\frac{\partial f}{\partial \tau}+\frac{\partial f}{\partial x^i}\frac{dx^i}{d\tau}+\frac{\partial f}{\partial q}\frac{dq}{d \tau}+\frac{\partial f}{\partial\hat q^i}\frac{d\hat q^i}{d \tau}=I_c\,,\label{eq1}
\end{alignat}
where $f\equiv f(x^i,\tau,q,\hat q^i) $ is the phase-space distribution and $I_c$ represents the collisions between particles. Here variables are: comoving coordinates of the particle $x^i$, conformal-time $\tau$ (\,$a(\tau)\,d\tau=dt$, $t$ = physical or proper time coordinate), comoving-momentum ${\bm q}$, and  $\bm{\hat  q}$ is the unit vector along $\bm q$. $\int d^3q\, I_c =0$, leads to the total conservation of phase-space distribution.

In the presence of anisotropies and inhomogeneities, the distribution function can be written as
\begin{alignat}{2}
f(\bm x,\tau, q,\hat{\bm q})=f_o(q,\tau)\,+ f_o(q,\tau)\ \Psi(\bm x, \tau, q,\hat{\bm q})\,,\label{eq2}
\end{alignat}
where back-ground distribution depends only on conformal-time and comoving-momentum amplitude, and $\Psi(\bm x, \tau,q,\hat{\bm q})$ is the first-order perturbation in phase-space distribution which depends on comoving spatial-coordinate, momentum, and conformal-time.  For the length-scale $\epsilon_{k}\gg1$, the DM consistent with the $\Lambda$CDM is nonrelativistic and noninteractive matter (CCDM), for which zeroth-order distribution function can be written as \cite{Dodelson:2003,Ma:1995E}
\begin{equation}
f_o(q,\tau)\propto \exp\left[-\frac{q^2}{2ma^2(\tau)T(\tau)}\right]\label{3}\,,
\end{equation}
where  $a(\tau)$ is the scale factor, $m$ is the mass of DM particles, and $T(\tau)$ is temperature of the CCDM scales as $a^{-2}(\tau)$. Hence $f_o(q,\tau)\equiv f_o(q)$ i.e. only depends on particle's comoving momentum ($\ q_i\equiv a(\tau)\,p_i$, where $|\bm p|\propto 1/a(\tau)$ is particle's physical momentum ). We have considered the line element in the conformal Newtonian gauge as \cite{MUKHANOV:1992HR,Ma:1995E}
\begin{equation}
ds^2=a^2(\tau)\left[-e^{2\psi}d\tau^2+e^{-2\phi}d\bm x^2\right]\,,\label{eq4}
\end{equation}
where $\psi\equiv\psi(\bm x,\tau)$ and $\phi\equiv\phi(\bm x,\tau)$ are scalar perturbations and corresponds to the Newtonian potential and perturbation to the spatial curvature (with a minus sign) respectively \cite{Dodelson:2003} . Since $\partial f/\partial \hat q^i$ and $d\hat q^i/d\tau$, both are the first-order quantity, we can neglect the last term of L.H.S up-to the first-order contribution in the equation \eqref{eq1}. For CCDM, the Boltzmann equation takes the form,
\begin{equation}
f_o\left[\frac{\partial \Psi}{\partial \tau}+\frac{\partial \Psi}{\partial x^i} \frac{q^i}{\epsilon}\right] + \frac{\partial f_o}{\partial q}\, \left[q\, \dot\phi-\epsilon\hat q^i\partial_i\psi\right]=0\,,\label{eq5}
\end{equation}
where $\epsilon\equiv\epsilon(q,\tau)=\sqrt{q^2+(am)^2}$ is the comoving energy of a particle \cite{Shoji:2010K}. Taking Fourier transformation of the linear perturbation $\Psi(\bm x, \tau,q,\hat{\bm q})$ and expanding in the form of Legendre polynomials $P_l$,
\begin{equation}
\Psi(\bm x, \tau,q,\hat{\bm q})=\sum_{l=0}^{\infty} (-i)^l\,(2l+1)\,\Psi_l(\bm k,\tau,q)\, P_l(\varsigma) \,,\label{eq6}
\end{equation}
where $\varsigma=\bm{\hat{k}}\cdot \bm{\hat{q}}$ , $\hat {\bm k}$ is the unit vector of $\bm k$ and $\Psi_l(\bm k,\tau,q)$ are coefficients of the Legendre polynomials. We get the differential equations for moments (or Boltzmann hierarchy),
\begin{alignat}{2}
\dot\Psi_0&=-kv_p\Psi_1-\dot\phi(k,\tau)\frac{d\ln f_o}{d\ln q}\,,\label{eq7}\\
\dot\Psi_1&=\frac{1}{3}kv_p\left[\Psi_0-2\Psi_2\right]-\frac{k}{3v_p}\psi(k,\tau)\frac{d\ln f_o}{d\ln q}\,,\label{eq8}\\
\dot\Psi_l&=kv_p\left[\frac{l}{2l+1}\Psi_{l-1}-\frac{l+1}{2l+1}\Psi_{l+1}\right]; \, \, \,l\geq 2\,,\label{eq9}
\end{alignat}
where $v_p=q/\epsilon$ is the particle's peculiar velocity. The time evolution of moments can be taken to the order of the Hubble time at long wavelength,
\begin{equation}
\Psi_l\sim (kv_p\mathcal{H}^{-1})^{l-2}\ \Psi_2\,; \, \, \, l \ge 2.\label{eq10}
\end{equation} 
Where $\mathcal{H}=a'/a$ and $a'=da/d\tau$. Thus, it is  clear that higher order moments can be written in terms of second order moment for $l>2$. If the factor of $\Psi_2$ in equation \eqref{eq10} is smaller than unity 
(i.e. $kv_p\mathcal{H}^{-1}\ll1$), then it implies the fluid approximation or truncation of the Boltzmann hierarchy. Taking a bound on the maximum possible particle velocity from the velocity in the non-linear regime \cite{Baumann:2010tm,Gramann:1998},
\begin{equation}
v_p^2\leq\Delta_v^2(k_{\rm NL}) \sim \Delta^2_\delta(k_{\rm NL}) \frac{\mathcal{H}^2}{k^2_{\rm NL}}\sim\frac{\mathcal{H}^2}{k^2_{\rm NL}}\,,\label{eq11}
\end{equation}
where, $\Delta_v^2(k)=({k^3}/{2\pi^2})\,P_v(k)$,  $P_v(k)=\langle|v_p(k)|^2\rangle$ is the power spectrum of velocity fluctuations and $\Delta_\delta^2 (k) = ({k^3} / {2\pi^2})\ P_\delta(k)\,$. $\Delta_\delta^2(k=k_{\rm NL})\sim 1$ corresponds to the separation between linear and non-linear scales. $P_\delta(k)=\langle|\delta(k)|^2\rangle$ is the power spectrum of density fluctuations.  Therefore,
\begin{equation}
kv_p\mathcal{H}^{-1}\lesssim \frac{k}{k_{\rm NL}}\,,\label{eq12}
\end{equation}
if $\epsilon_{k}\gg 1$ then $kv_p\mathcal{H}^{-1}\ll1$, implies \textit{fluid approximation}
(i.e. $l_{max}=2$). Thus, for linear scale $\epsilon_{k} \gg 1$, the higher moments are suppressed (i.e. $\Psi_l\ll\Psi_2\,$ for $l>2\,$). $\Psi_1$ and $\Psi_2$ give energy flux and shear stress respectively. This hierarchy depends on non-linear scale and it comes due to the gravitational coupling of fluid.


\section{Relaxation time for CCDM and SIDM}\label{sect:3}
In above section [\ref{sect:2}] we have obtained the fluid approximation for CCDM. In this section we will calculate the mean free-time (relaxation time) for collisionless cold dark-matter and self-interacting dark-matter.


\subsection{Collisionless cold dark-matter} Taking that, in a Hubble time CCDM particle move to the scale $v_p\,\tau_{cb}$, one can rewrite  inequality \eqref{eq12}  as
\begin{equation}
kv_p\tau_{cb}\lesssim \frac{k}{k_{\rm NL}}\,,\label{eq13}
\end{equation}
Multiplying equation \eqref{eq13} by $f_o(q)$, where $q=v_p\epsilon\cong v_p\,a\,m$, and  taking integral over $d^3q$
\begin{equation}
\frac{1}{am}\tau_{cb}\int d^3q\,q\,f_o(q)\lesssim \frac{1}{k_{\rm NL}}\int d^3q\,f_o(q) \,,\label{eq14}
\end{equation}
\begin{equation}
\tau_{cb} \,\bar v_p\lesssim \frac{1} { k_{\rm NL}}\,,\label{eq15}
\end{equation}
here $\bar v_p\cong{\bar q}/{(a\,m)}\,$ is the mean peculiar velocity of fluid, where $n(x,\tau)\,\bar q=1/a^3\,\int d^3q\,q\,f_o(q)$; $n(x,\tau)=1/a^3\int d^3q\, f_o(q)$ is the number density and $\bar{v}_p\tau_{cb}$ is regarded as the ``mean free path". Therefore from equation \eqref{eq15} , we write
\begin{equation}
\tau_{cb}^{-1}\gtrsim \,\bar v_p\, k_{\rm NL}\,.\label{eq16}
\end{equation}
Here, the relaxation time arises because particles are gravitationally bound and during a Hubble time particles move only up-to the nonlinear scale. In the absence of gravitational coupling or non-linear scale ($\,k_{\rm NL}\,$), the mean free-path can be infinitely long. Here we would like to note that, $k^{-1}_{NL}$ refers to the objects of galaxy clusters size. $k^{-1}_{\rm NL}$ can be estimated by considering $\Delta_\delta^2(k_{\rm NL})=1\,$ \cite{Kolchin:2009,Widrow:2009}. For redshift $z=0$, we get $k_{\rm NL}\approx 0.2$~{h/Mpc} \cite{Kolchin:2009,Galaxy:2019,Jenkins:1998}.

\subsection{Self-interacting dark-matter} 
In the above subsection,  we have obtained the relaxation time $\tau_{cb}$ for CCDM in the presence of nonlinearities. For the case of cold collisionless dark-matter, relaxation time arises because of the nonlinear structures due to self-gravity. For the case of self-interacting dark-matter, the concept of mean free path arises due to collisions between particles. But for the present case, we need to consider the effects of self-gravity and self-interaction. Thus our formalism involves relaxation times due to both these effects. For the case of SIDM \cite{Spergel:2000DSP, Tulin:2017Y, Kaplinghat:2016TY}, relaxation time can be written as \cite{Hannestad:2000S,Jacoboni:2010},
\begin{equation}
\tau_{si}^{-1}=n\,\langle\sigma\, v_c\rangle\,,\label{eq17}
\end{equation}
where $\langle\cdot\cdot\rangle$ represents the ensemble average, $n$ is the number density of the particles, $\sigma$ is the differential cross-section for scattering and $v_c=|\bm{\mathit{v}}_c|$ is the relative velocity between DM particles.


\section{Spatial perturbation in the MB distribution of DM }\label{sect:4}
In the present case, the relaxation time comes from two different processes, one from the gravitational coupling of the DM particle's and other one from DM self interaction. The collision term ($I_c$) in equation \eqref{eq1}, can be approximated by ``relaxation time approximation''. Thus for the present case, the collision term $I_c$ becomes \cite{Chapman:1952, Groot:1969P, ANDERSON:1974H, Kerson:1987, Hannestad:2000S} 
\begin{equation}
I_c\approx-\frac{f-f_o}{\tau_{cb}}-\frac{f-f_o}{\tau_{si}}=-\frac{f-f_o}{\tau_\mathit{eff}}\ ,\label{eq18}
\end{equation}
where $\tau_\mathit{eff}^{-1}=\tau_{cb}^{-1}+\tau_{si}^{-1}$ and $f=f(\bm x,\tau,\bm q)$ are the inverse effective relaxation time and phase-space distribution function respectively. At the lowest order approximation, we can assume $f_o$ to be the Maxwell-Boltzmann distribution \cite{Atreya:2018bm},
\begin{alignat}{2}
f_o(\bm x,\tau,\bm q)=\frac{g }{(2\pi)^3}&\, \exp\left[\frac{P_\mu U^\mu}{T}\right]=\frac{g }{(2\pi)^3}\nonumber\\
\times &\exp\left[-\frac{\epsilon(\tau,q)}{a(\tau)\,T(\tau)}+\frac{\bm{q}\cdot\bm{V(\tau,\bm{x})}}{a(\tau)T(\tau)}\right]\,,\label{eq19}
\end{alignat}
Where we have used the metric \eqref{eq4}. Here four velocity $U^\mu$ satisfies $U^\mu U_\mu=-1$ and $P_\mu U^\mu=P_0U^0+P_iU^i$, where $U^0=a^{-1}(\tau)\,e^{-\psi(\tau,\bm x)}$, $P_0=-\epsilon\,e^{\psi(\tau,\bm x)}$, $U^i=a^{-1}(\tau)V^i$ with $V^i\equiv dx^i/d\tau$ is the coordinate velocity of the fluid. $P_i$ is replaced by $q_i$, and $\hat q^i\hat q_i=\delta_{ij}\hat q^i\hat q^j=1$, where $\bm{\hat  q}$ is the unit vector along $\bm q$, as in the references \cite{Ma:1995E,Baumann:2010tm}. In equation \eqref{eq19},  $\bm q\cdot\bm V(\tau,\bm x)=\delta_{ij}\,q^i\,V^j(\tau,\bm x)\,$ and $g$ represents the degree-of-freedom. 
\noindent
Writing $f(\bm x,\tau,\bm q)=f_o(\bm x,\tau,\bm q)+\delta_f(\bm x,\tau,\bm q)$, where $\delta_f(\bm x,\tau,\bm q)$ is the variation from the MB distribution. The Boltzmann equation in this case, takes the form 
\begin{equation}
\left[\frac{\partial }{\partial \tau}+\frac{dx^i}{d\tau}\frac{\partial }{\partial x^i}+\frac{dq}{d \tau}\frac{\partial }{\partial q}\right](f_o+\delta_f)=-\frac{\delta_f}{\tau_\mathit{eff}}\ .\label{eq20}
\end{equation}
Assuming $\delta_f\ll f_o$, we can neglect $\delta_f$ on the L.H.S., implies 
\begin{equation}
\delta_f=-\tau_\mathit{eff}\left[\frac{\partial }{\partial \tau}+\frac{dx^i}{d\tau}\frac{\partial }{\partial x^i}+\frac{dq}{d \tau}\frac{\partial }{\partial q}\right]f_o\,.\label{eq21}
\end{equation} 
Obtained $\delta_f\equiv\delta_f(\bm x,\tau,\bm q)$ depends on the effective relaxation time of the fluid. In the above equation, first term is related with the heat conduction \cite{Weinberg:1972}. Second term defines the spatial changes in the fluid with velocity i.e. related with spatial-dissipation in the fluid. In the third term, conformal time-derivative of comoving-momentum $q$ can be written in terms of the conformal time-derivative and comoving spatial-derivative of the scalar perturbations $\phi$ and $\psi$ respectively ($dq/d\tau=q\dot\phi-\epsilon\hat q^i\partial_i\psi$). This term signifies effect of the  over/under-dense regimes or fluctuations in the phase-space distribution of the DM particles. Viscosity in the fluid  is defined by the spatial derivative of the fluid velocity, and in the distribution function ($\,f_o(\bm x,\tau,\bm q)\,$), only fluid velocity depends on spatial component. Accordingly we evaluate only the second term of the equation \eqref{eq21},
\begin{alignat}{2}
\delta_\mathit{fs}=-\tau_{\mathit{eff}}\,\frac{1}{a\,T\,\epsilon}\,\Bigg[\,q^iq^l\,\bigg\{\,\frac{1}{2}\,\big(\,\partial_i(V_l)+\partial_l&  (V_i)\,\big)
-\frac{1}{3} \delta_{il}\, \theta \bigg\}\nonumber\\
&+\frac{1}{3}\,q^2\,\theta\,\Bigg] \ f_o\label{eq22}\,,
\end{alignat} 
where $\delta_\mathit{fs}\equiv \delta_\mathit{fs}(\bm x,\tau,\bm q)$ is the spatial first order perturbation in the phase-space distribution  , $\theta=\partial_jV^j$ is the velocity divergence and it's related with the bulk-viscosity. The quantity, in the curly bracket, is known as the shear tensor $\sigma_{il}$ \cite{ Baumann:2010tm,Weinberg:1971,Weinberg:1972}.


\section{Viscosity in the dark-matter fluid}\label{sect:5}
The stress-energy tensor for imperfect fluid can be written as \cite{Baumann:2010tm},
\begin{equation}
T_{ij}=\rho U_iU_j+(p-\zeta\theta)\gamma_{ij}+\Sigma_{ij}\,,\label{eq23}
\end{equation}
where $\rho$ is the energy density, $p$ is the pressure, $\bm U$ is the fluid velocity, $\zeta$ is the bulk viscosity, $\Sigma^{ij}$ is the viscous stress-tensor, $\gamma^{ij}=\textsf{g}^{ij}+U^iU^j$ and $\textsf{g}^{ij}$ is metric. Here, we are interested in the bulk-viscosity and shear-viscosity as the dissipation in DM fluid. The viscous stress-tensor defined as \cite{Baumann:2010tm},
\begin{equation}
\Sigma_{ij}=-\eta \sigma_{ij}\,,\label{eq24}
\end{equation}
where $\eta$ is shear viscosity. Thus the dissipative stress-energy tensor 
\begin{equation}
\Delta T_{ij}\cong-\eta\, \sigma_{ij}-\zeta\, \theta\, \delta_{ij}\,.\label{eq25}
\end{equation}
The stress-energy tensor can be described in the terms of the distribution function \cite{Ma:1995E}
\begin{equation}
\overline T_{ij}+\Delta T_{ij}= \frac{1}{a^4}\int q^2 dq d\Omega\frac{{q}_i{q}_j}{\epsilon(q,\tau)}(f_o+\delta_\mathit{fs})\,,\label{eq26}
\end{equation}
here we are interested only in spatial dissipation, therefore we have taken only $\delta_\mathit{fs}$, and $\overline T_{ij}$ is the background energy-momentum tensor. Substituting equation \eqref{eq22} into equation \eqref{eq26} and comparing with equation \eqref{eq25}, we get the expression for the effective bulk viscosity as
\begin{equation}
\zeta_\mathit{eff}=\frac{1}{9}\cdot\tau_\mathit{eff}\cdot\frac{1}{a\,T}\cdot\frac{1}{a^4}\int d^3q\,\frac{q^4}{\epsilon^2}f_o\,,\label{eq27}
\end{equation}
and for the effective shear viscosity as \cite{Sean:1985,Kadam2015},
\begin{equation}
\eta_\mathit{eff}=\frac{1}{15}\cdot\tau_\mathit{eff}\cdot\frac{1}{a\,T}\cdot\frac{1}{a^4}\int d^3q\,\frac{q^4}{\epsilon^2}f_o\,.\label{eq28}
\end{equation}
For the cold (non-relativistic) DM, the comoving energy ($\epsilon$) can be approximated as $\epsilon^{-2}\simeq (am)^{-2}-q^2/(a^2m^2)^2$. Hence
\begin{equation}
\zeta_\mathit{eff}\cong\frac{1}{9}\cdot\tau_\mathit{eff}\cdot\frac{n}{m^2\, T}\cdot\frac{1}{a^4}\ \left[\langle\, q^4\,\rangle-\frac{1}{(am)^2}\,\langle \,q^6\,\rangle\right]\,,\label{eq29}
\end{equation}
and
\begin{equation}
\eta_\mathit{eff}\cong\frac{1}{15}\cdot\tau_\mathit{eff}\cdot\frac{n}{m^2\, T}\cdot\frac{1}{a^4}\ \left[\langle\, q^4\,\rangle-\frac{1}{(am)^2}\,\langle \,q^6\,\rangle\right]\,,\label{eq30}
\end{equation}
where $n(\bm x,\tau)\,\langle\, q^4\,\rangle=a^{-3}\int d^3q\,q^4\,f_o(\bm x,\tau,\bm q)\,$, $n(\bm x,\tau)$ is the number density of the DM and   $\rho(\bm x,\tau)=m\,n(\bm x,\tau)$ is the energy density of the DM. We can write equation \eqref{eq29} and \eqref{eq30} as
\begin{equation}
\frac{1}{\zeta_\mathit{eff}}=\frac{1}{\zeta_\mathit{SIDM}}+\frac{1}{\zeta_\mathit{CCDM}}\,,\label{eq31}
\end{equation}
and 
\begin{equation}
\frac{1}{\eta_\mathit{eff}}=\frac{1}{\eta_\mathit{SIDM}}+\frac{1}{\eta_\mathit{CCDM}}\,.\label{eq32}
\end{equation}
Where $\zeta_\mathit{SIDM}$ and $\zeta_\mathit{CCDM}$ are bulk-viscosities due to self-interacting DM and gravitational coupling of DM respectively, and defined as
\begin{alignat}{2}
{\zeta_\mathit{SIDM}}=\frac{1}{9}\,{\tau_\mathit{si}}\,{S}\qquad  \ \text{and}\qquad  \ {\zeta_\mathit{CCDM}}=\frac{1}{9}\,{\tau_\mathit{cb}}\,{S}\,,\label{eq33}
\end{alignat}
similarly, $\eta_\mathit{SIDM}$ and $\eta_\mathit{CCDM}$ are shear-viscosities due to SIDM and gravitational coupling of DM respectively, and defined as
\begin{alignat}{2}
{\eta_\mathit{SIDM}}=\frac{1}{15}\,{\tau_\mathit{si}}\,{S}\qquad  \ \text{and}\qquad  \ {\eta_\mathit{CCDM}}=\frac{1}{15}\,{\tau_\mathit{cb}}\,{S}\,,\label{eq34}
\end{alignat}
here $S\equiv S(\bm x, \tau)$, 
\begin{equation}
S=\frac{n}{m^2\,T}\,\frac{1}{a^4}\left[\langle\, q^4\,\rangle-\frac{1}{(am)^2}\,\langle \,q^6\,\rangle\right]\label{eq35}\,.
\end{equation}
We get the effective bulk-viscosity \eqref{eq31} and shear-viscosity \eqref{eq32} due to two different relaxation times because of two different processes, as in the reference \cite{Asakawa:2007sb}, and these are inversely additive. 


\section{Result and Discussion}\label{sect:6}
In the present work, we have considered the possibility where the viscosity coefficients of a dark-matter fluid can arise due to two different processes. For this purpose, we have used the Boltzmann equation with the effective relaxation time \eqref{eq18}, which
contains contributions from the nonlinear scale and the self-interaction between the dark-matter particles. Here we note that the relaxation times for the different processes in the Boltzmann equation are inversely additive. This leads to the expressions of the effective bulk \eqref{eq31} and shear \eqref{eq32} viscosities. In terms of relaxation time one can write  the effective (shear or bulk)  viscosity $\eta_\mathit{eff}\ \text{or}\ \zeta_\mathit{eff}\propto \, (\,\tau_{cb}\tau_{si}\,)/(\,\tau_{cb}+\tau_{si}\, )$. Thus the shorter relaxation time is dominated in determining the viscous contribution. For example, when relaxation time due to the self-interaction is larger than the relaxation time due to the nonlinearities, the effective viscosity is dominated by the smaller time scale i.e. nonlinearities. 


Now, for example, consider the relaxation time arising due to self-interaction  $\tau_{si}=m/(\rho\, \langle \sigma v_c \rangle)$ and the  constraints on $\sigma/m$ discussed in Ref. \cite{Kaplinghat:2016TY}. As argued before $\tau_{si}$ and the age of the halo are related $\tau_{si}\sim t_{age}$. Thus, one gets  $\tau_{si} \sim 3.16\times 10^{16}$~sec for a super cluster, $\tau_{si}\sim 1.58\times 10^{17} $~sec for cluster and $\tau_{si} \sim 3.16\times 10^{17}$~sec for galaxy scales. Next consider the relaxation time $\tau_{cb}$ for the cold-collision less case in the effective fluid theory \cite{Baumann:2010tm}. Expression for the relaxation time: $\tau_{cb}\lesssim \, k_{\rm NL}^{-1}/\bar v_p$, where $v_p$ is particle velocity in the nonlinear regime and $k_{\rm NL}\approx 0.2 $~h/Mpc, can be estimated by using the relation $\Delta_\delta^2(k_{\rm NL})=1\,$ for $z=0$ \cite{Kolchin:2009,Widrow:2009,Galaxy:2019,Jenkins:1998}. Three-dimensional root-mean-square peculiar velocity of matter smoothed over a  radius $3~{\rm h^{-1}Mpc}$ has been estimated to $507\pm48$~Km/sec \cite{Bahcall:1996,Gramann:1998}. We take the $h=0.70$ \cite{Planck:2018}. Thus we get  $ \tau_{cb}\lesssim \, 3.97\times 10^{17}$~sec. From equation \eqref{eq35}, we get $S=2.1\times10^{-15}$~Kg /meter /sec$^2$ using Equipartition of energy $\bar v_p=\sqrt{(3T/m)}$ for $a=1$. We estimate shear and bulk viscosity coefficients to be $\lesssim 5.6\times10^1$~Kg /meter /sec and $\lesssim 9.3\times10^1$~Kg /meter /sec respectively for collisionless cold dark-matter case for redshift $z=0$. For the SIDM case, we get shear and bulk viscosity coefficients to be $2.2\times10^1$~Kg /meter /sec and $3.7\times10^1$~Kg /meter /sec respectively for $a=1$ and cluster scales. Therefore, effective shear viscosity $\eta\lesssim1.6\times10^1$~Kg /meter /sec and bulk viscosity $\zeta\lesssim2.6\times 10^1$~Kg /meter /sec for cluster scale and $z=0$.  Authors of the Ref. \cite{ Bhatt:2019mn}, consider model dependent bulk viscosity in the light of Experiment to Detect the Global Epoch of Reionization Signature (EDGES) observation and constraint bulk viscosity coefficient $\zeta\lesssim3.93$~ Kg /meter /sec for constant viscosity case--no dependency on redshift. For variable viscosity case $\zeta\lesssim1.57\times10^4$~Kg /meter /sec. In the dimensionless form (multiplying by $8\pi G/H_0=8.2\times 10^{-9}$~meter-sec /Kg) shear and bulk viscosity coefficients are $\sim1.8\times10^{-7}$ and $\sim3\times10^{-7}$ respectively for SIDM case for the cluster scale. For the super cluster scale, shear and bulk viscosities in the dimensionless form are $\sim 3.6\times10^{-8}$ and $\sim 6\times10^{-8}$ respectively. For CCDM case, dimensionless shear and bulk viscosity are $\lesssim 4.6\times10^{-7}$ and $\lesssim 7.6\times10^{-7}$ respectively. To reduce discordance between PLANCK and LSS data, author of the Ref. \cite{Anand:2017} consider viscous dark-matter without self interaction between DM particles. Authors found that, dimensionless $ 2.0\times 10^{-7} \leq \eta \leq  2.2\times10^{-6}$ and $3.2 \times 10^{-7} \leq \zeta \leq 3.32 \times 10^{-6}$. These values of shear and bulk viscosities are consistent with the effective fluid description transport coefficients based on CCDM \cite{Baumann:2010tm}. While, for SIDM case shear and bulk viscosity are $\mathcal{O}(10^{-1})$ small. But, for small scales these coefficients are consistent with SIDM case also. In the Ref. \cite{Velten:2013}, authors consider model-dependent bulk viscosity: $\zeta=\zeta_0(\rho/\rho_0)^\nu$. Here, $\rho\equiv\rho(\tau)$ is dark-matter energy density and $\rho_0=\rho(z=0)$. They get upper constraint on constant bulk viscosity coefficient ($\nu=0$) to the $\lesssim1.95\times 10^2$~Kg /meter /sec for $k=0.2$~h/Mpc by requiring that perturbations should grow to the nonlinear stage. The authors also discuss the upper constraint on $\zeta$ for variable bulk viscosity ($\nu\neq0$). As we have shown, shorter relaxation time contributes more to the viscosity because the total viscosity of the system depends on $\tau_\mathit{eff}$.  In this work, we have considered the viscosity of the cosmic fluid at the cluster and shown that the effective viscosity of the fluid can reduce by a factor of $\sim$2.  

In conclusion, we have examined the role of viscosity  due to self-interaction. It is shown that such viscosity should not be considered in isolation as  since the effective theory description based on $\Lambda$CDM model also has viscosity and both the viscous coefficients are added inversely.  From the examples we have considered above, at the cluster scale the effective fluid  description  of $\Lambda$CDM models provide good estimates of viscosity. 


\end{document}